\documentclass[11pt]{iopart}
\usepackage{iopams}  
\usepackage{braket}
\usepackage{graphicx}
\usepackage[colorlinks=true,linkcolor=blue,citecolor=blue,urlcolor=blue]{hyperref}
\begin{document}

\title[Polyatomic Molecules as Quantum Sensors for Fundamental Physics]{Polyatomic Molecules as Quantum Sensors for Fundamental Physics}

\author{Nicholas R. Hutzler}

\address{California Institute of Technology \\ Division of Physics, Mathematics, and Astronomy \\ Pasadena, CA 91125}
\ead{hutzler@caltech.edu}
\vspace{10pt}
\begin{indented}
\item[]August 2020
\end{indented}

\begin{abstract}
Precision measurements in molecules have advanced rapidly in recent years through developments in techniques to cool, trap, and control.  The complexity of molecules makes them a challenge to study, but also offers opportunities for enhanced sensitivity to many interesting effects.  Polyatomic molecules offer additional complexity compared to diatomic molecules, yet are still ``simple'' enough to be laser-cooled and controlled.  While laser cooling molecules is still a research frontier itself, there are many proposed and ongoing experiments seeking to combine the advanced control enabled by ultracold temperatures with the intrinsic sensitivity of molecules.  In this perspective, we discuss some applications where laser-cooled polyatomic molecules may offer advantages for precision measurements of fundamental physics, both within and beyond the Standard Model.

\end{abstract}


%
%
%
%
\ioptwocol

\newcommand{\E}{\mathcal{E}}
\newcommand{\Eeff}{\mathcal{E}_{ef\!f}}

\section{Introduction}

Atoms and molecules can serve as sensitive probes of essentially every particle, force, and interaction -- both known and unknown~\cite{Safronova2018}.  From measuring gravity to searching for supersymmetry, atoms and molecules make profound contributions to our understanding of the universe, often using experiments that can fit on a (possibly large) table.  An important advantage is the availability of techniques to control their degrees of freedom, both classical and quantum, and make measurements with extreme precision.  In this perspective we will focus on an emerging area -- laser-cooled polyatomic molecules -- and the contributions which they could make to fundamental physics.

Molecules are of interest not just for their applications to fundamental physics~\cite{Safronova2018,DeMille2017,Chupp2019,Cairncross2019}, but also to quantum chemistry~\cite{Balakrishnan2016,Bohn2017}, quantum matter~\cite{Bohn2017,Baranov2012}, quantum information~\cite{DeMille2002}, and more.  Their complexity is a powerful feature; they have more states, more energy scales, more interactions, and more interacting bodies compared to atoms.  Complexity is not always better, as it makes them difficult control and understand, but advances in experimental and theoretical techniques have started to make possible many of the exciting promised applications.  Advanced methods to cool and control molecules, including laser cooling~\cite{Tarbutt2018,McCarron2018}, promise a revolution in molecular science like they did for atomic science.

The additional internal degrees of freedom and chemical diversity of polyatomic molecules make them a promising platform for many areas of research.  In this perspective we will focus on fundamental physics, but polyatomic molecules also have specific advantages for quantum information and many-body physics~\cite{Wei2011,Wall2013,Wall2015,Yu2019}, and give many more chemical bonds and reactions to study with full quantum control~\cite{Balakrishnan2016}.  One critical difference compared to diatomic molecules~\cite{DeMille2015} is that polyatomics, by which we mean molecules with three or more atoms, have additional \emph{mechanical} degrees of freedom -- they can rotate and vibrate in many ways.  How these additional degrees of freedom can benefit precision measurements is the focus of this perspective.

First, we begin with a very brief introduction to the relevant properties of polyatomic molecules, including their multiple degrees of freedom, and the ability to generically polarize them in small electric fields.  Next, we discuss some aspects of laser cooling molecules, and how these techniques are applicable to polyatomic species.  Finally, we describe some searches for fundamental physics which could benefit from the unique features of polyatomic molecules, in particular those which can be laser-cooled.

\section{Polyatomic Molecules}

The structure of polyatomic molecules is fantastically complex and diverse, so here we will only describe the features critical to understanding the experiments that we discuss~\cite{Townes2012,Herzberg1967,Bernath2005}.   A discussion of some important features of diatomic molecules can be found in the appendix.  The first main difference relative to diatomics is that polyatomic molecules have multiple vibrational degrees of freedom.  The vibrational state is often specified as $(\nu_1\ldots\nu_n)$, where $\nu_i$ is an integer counting the number of quanta in the $i^{th}$ mode.  For example, a linear triatomic of the form MOH, where M is much heavier than O, has three modes: M--O stretch, M--O--H bend, O--H stretch, which are typically denoted as modes 1, 2, 3, respectively.

Bending modes of a linear bond are doubly-degenerate since the bending can occur in either of two perpendicular planes.  Superpositions of these two modes with a phase difference correspond to states where the middle nucleus is orbiting around the symmetry axis with angular momentum $\ell$, as though it was a permanently bent molecule rotating around its symmetry axis.  States with well-defined $\ell$ are not parity eigenstates since reflection will interchange $+\ell\leftrightarrow-\ell$, so in free space the eigenstates are equal superpositions $\ket{+\ell}\pm\ket{-\ell}$.  These states have opposite parity and are split by an amount called the $\ell-$doubling, typically tens of MHz.  This is similar in many respects to $\Lambda$-doubling (see the appendix), and we shall consider it in more detail later.  Therefore, each rotational level in these vibrational states is split into two states of opposite parity, as opposed to a ladder of rotational states with a well-defined, alternating parity when these doublets are absent. States which can support vibrational angular momentum include a superscript $\ell$ in their state labels, such as $(\nu_1\nu_2^\ell\nu_3)$ for a linear triatomic (though this is often omitted for $\nu_2=1$, as there is only one possible value of $\ell$).

An important type of non-linear molecule is the symmetric top, which has three non-zero moments of inertia but with two of them equal (unlike a linear molecule, which has two moments of inertia equal and one vanishing) such as NH$_3$ or CaOCH$_3$.  These molecules have additional rotational structure due to rotation about the symmetry axis, quantized by the number $K$.  States with $\pm K$ rotational quanta about the symmetry axis are nominally degenerate, but like $\ell$-doubling the field-free eigenstates are even and odd superpositions of $\pm K$ with opposite parity.  These $K$-doublets are split by an amount even smaller than $\ell-$doublets, typically $\lesssim$1~MHz for the types of molecules considered here~\cite{Klemperer1993,Butcher1993}.  Asymmetric top molecules, which have three distinct moment of inertia, are also of interest and all of the advantages we shall discuss generally apply to them as well~\cite{Augenbraun2020ATM}, though we will not consider them specifically here.

\subsection{Parity doubling and polarization}

Parity doublets are a useful resource since they enable polarization in small electric fields.  The electric dipole operator $\vec{d}=e\vec{r}$ is parity-odd, and any parity eigenstate $\ket{\psi}$ must therefore have a dipole moment of zero: $\braket{\psi | \vec{d} | \psi}=0$.  Since parity is a symmetry of free space (to good approximation), atomic and molecular eigenstates are parity eigenstates in the absence of applied fields.  Non-zero dipole moments therefore arise as a result of mixing states of opposite parity, for example by applying a static electric field $\vec{\E}$.  The small splitting of parity doublets makes this happen at lower fields.

\begin{figure}[ht]
\centering
\includegraphics[width=0.48\textwidth]{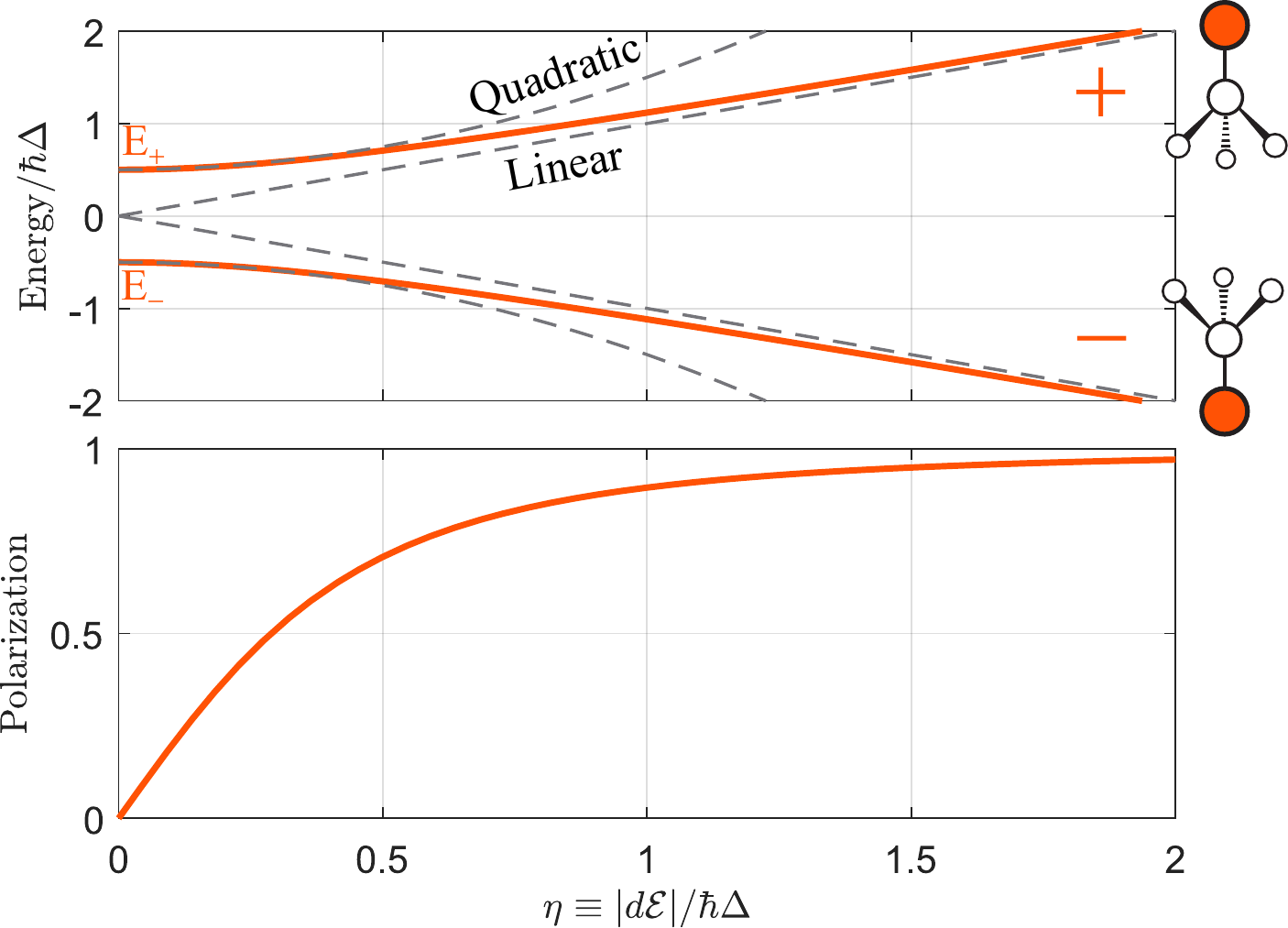}
\caption{Stark shifts and polarization in a model two level system.  \emph{Top}: The actual shifts (thick solid lines) transition from quadratic at low field to linear at high field, with a cross-over point where $|d\E|\approx\hbar\Delta$.  The states with linear shifts correspond to the molecular dipole polarized in the lab frame in one of two directions.  This pair of states with equal-and-opposite polarization is unique to systems with parity doublets. \emph{Bottom:} Polarization is linear at low fields as the molecules have an induced dipole, and saturates at high fields as the dipole becomes fully induced.}
\label{fig:polarizationPlots}
\end{figure}

As an example, consider a two level system split by energy $\hbar\Delta$ with states coupled by an electric field $\E$,
\[ H = \left(\begin{array}{cc} -\hbar\Delta/2 & -d\E \\ -d\E & +\hbar\Delta/2 \end{array}\right). \]
The energies of this system are given by
\[ E_\pm = \pm\frac{1}{2}\hbar\Delta\left[1+4\eta^2\right]^{1/2}, \]
where we have defined $\eta\equiv |d\E|/\hbar\Delta$.  At very high fields, $\eta\gg 1$, we obtain the usual linear Stark shift $\E_\pm\approx\pm d\E$, whereas at low field we obtain quadratic shifts $\Delta\E_\pm\approx d^2\E^2/\hbar\Delta$, as shown in figure \ref{fig:polarizationPlots}.  This is the characteristic behavior of an induced dipole, where the degree of polarization $P$ is

\[ P = \left|\frac{\partial E_\pm}{\partial(d\E)}\right| = \frac{2\eta}{\sqrt{1+4\eta^2}}.\]

The scale of the cross-over point from linear to quadratic is $\eta\approx 1$, or $\hbar\Delta\approx d\E$.  Therefore, the field required to polarize something reduces linearly with the splitting of opposite parity states.  A typical, atomic unit of dipole moment $ea_0$ translates to an energy shift of $\approx$1~MHz in 1 V/cm of electric field.  Atoms must mix opposite parity electronic states, and are therefore limited to $P\approx\eta\approx 10^{-3}$ even in large fields of $\sim$100~kV/cm, whereas molecules can achieve $P\sim 1$ by mixing rotational states in fields of $\sim$10~kV/cm, or by mixing parity doublets in fields of $\lesssim 1$~kV/cm (even as low as ~1~V/cm).  Besides becoming polarized at smaller fields, the availability of states with equal-yet-opposite polarization has important implications which will be explored later.

Unfortunately, this splitting $\Delta$ is never zero~\cite{Klemperer1993}, so molecules always have induced dipole moments.  The sources of doubling mentioned above ($\ell$ and $K$), along with the $\Lambda$ or $\Omega$ doubling present in molecules with non-zero electronic orbital angular momentum, as well as other sources of doubling such as tunneling (the classic example being NH$_3$~\cite{Townes2012}), are always split by some small amount.  The specific origin is different for every source of doubling, but often involves the rotation of the molecule.  For example, the non-linearity in a bending mode reduces the symmetry of the molecule as it can be viewed as a bent molecule rotating around its symmetry axis.  This lifts the degeneracy of the moments of inertia about axes perpendicular to the symmetry axis, resulting in slightly different energy for the nominally degenerate bending modes as the molecule rotates.

\begin{figure}[ht]
\centering
\includegraphics[width=0.4\textwidth]{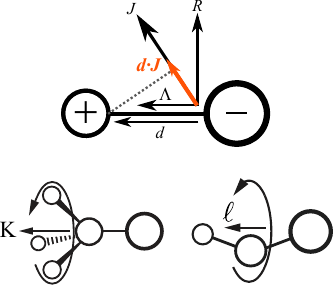}
\caption{The importance of angular momenta with non-zero projection on the internuclear axis $\hat{n}$.  \emph{Top}: For a molecule (shown as diatomic for simplicity) where the total angular momentum $\vec{J}$ has non-zero projection on the internuclear axis, $\vec{J}\cdot\hat{n}\neq 0$, the projection of the molecule-frame dipole moment on $\vec{J}$ is non-zero, $\vec{J}\cdot\vec{d}\neq 0$.  This occurs whenever there is some angular momentum quantized along $\hat{n}$, which can arise from electronic or mechanical means.  Since $\vec{J}$ is a good vector in the lab frame, the result is a linear Stark shift. \emph{Bottom}:  Angular momentum along the symmetry axis can , for example, due to rigid body rotations (left) bending (right).}
\label{fig:angMomOnAxis}
\end{figure}

There is a more qualitative argument to understand why states with $\Lambda,\ell$, and $K$ doublets can be easily polarized, and arises from the fact that they all are non-zero angular momentum projections on the internuclear axis $\hat{n}$, along which the molecular dipole points~\cite{Herzberg1989}.  Consider a molecule with $\ell>0$ (the cases for $\Lambda,K$ are similar) and rotation $\vec{R}\perp\hat{n}$, as shown in figure~\ref{fig:angMomOnAxis}.  These angular momenta are coupled via the rotational interaction described in the previous paragraph, or more generally Coriolis-type forces, so the total angular momentum $\vec{J}=\vec{\ell}+\vec{R}$ is the only good vector in the lab frame.  Accordingly, since the molecular dipole moment $\vec{d}\propto\hat{n}\propto\vec{\ell},$ we have that $\vec{d}\cdot\vec{J}\neq0.$  The non-zero projection of the dipole moment along a good, lab-frame quantum number means that there is a linear dipole moment in the lab frame, and in fact this projection argument even gives the correct magnitude of the effect.  Of course this argument ignores the fact that the $\pm\ell$ states are split, but intuitively captures the importance of angular momentum projections on the internuclear axis.  States with no angular momentum projection along the internuclear axis, for example $^2\Sigma$ linear molecules, correspondingly do not have linear Stark shifts, and even when fully polarized do not possess pairs of states with equal and opposite polarization.  

Parity doublets are already of importance in diatomic molecular precision measurements~\cite{Bickman2009,Cairncross2017,ACME2018}, where they arise as a result of $\Lambda$ or $\Omega$ doubling.  However, the generic availability of parity doublets in polyatomic molecules is largely independent of electronic structure.  Much of what we discuss here will depend on some particular electronic structure, such as laser cooling and sensitivity to fundamental symmetry violations, so polyatomic molecules offer new opportunities to combine these features.

As an aside, it is important to note that no permanent dipole moment has \emph{ever} been observed, and could only arise from exotic symmetry-violating physics as we shall discuss later.  Objects frequently claimed to have permanent electric dipole moments, such as H$_2$O, NH$_3$, NaCl, etc. in fact have \emph{induced} dipole moments.  These polar molecules transition from the quadratic to the linear Stark regime in ``very small'' fields, but have quadratic Stark shifts at sufficiently small fields indicative of induced dipoles.  The earlier discussion about parity proves that this must always be the case, but the physical chemistry behind why this happens in molecules is also understood~\cite{Klemperer1993,Butcher1993}.  The confusion is compounded by the fact that the quantity $d$ is usually called the permanent dipole moment of the molecule, since it can be viewed as the dipole moment in the molecular frame.  However, the molecular frame is a mathematical construct (albeit a physically intuitive one) and only the dipole moment in the lab frame is physically observable, and must be aligned by application of fields.

\section{Laser Cooling}

Laser cooling molecules has advanced with tremendous speed, from the first demonstration in 2010~\cite{Shuman2010} to the implementation of advanced cooling and trapping techniques in recent years.  Many of these developments are the topic of recent reviews~\cite{Tarbutt2018,McCarron2018}, and we will not repeat this very interesting story or summarize the results of the field here, apart from those directly relevant to our topics.  Instead, we will briefly review some of the key features and challenges of laser cooling diatomic and polyatomic molecules.

The motivation for laser cooling and trapping in precision measurements is extremely strong.  Ultracold and trapped species can offer long coherence times, large numbers, precise control of electromagnetic environments, and the opportunity for advanced quantum control and engineering~\cite{Hunter2012,Tarbutt2013,Kozyryev2017PolyEDM}.  The potential gain in sensitivity is orders of magnitude for many interesting sources of new physics, including those that we will consider in the next sections.  For example, searches for fundamental symmetry violation utilizing all of these advantages could probe energy scales far outside the reach of any particle accelerators, and several are already underway~\cite{Kozyryev2017PolyEDM,Parker2015,Norrgard2017,Tang2018,Lim2018,Aggarwal2018}.  We will consider laser-cooled neutral molecules here, though techniques with trapped molecular ions offer another route to major improvements without relying on laser cooling~\cite{Cairncross2017,Zhou2020}.

The basic idea of laser cooling molecules is identical to that of atoms~\cite{Metcalf1999}, and in its most common form relies on repeatedly exciting an electronic transition with a laser so that photons are scattered from the molecule.  This ``photon cycling'' transfers momentum to the molecule, and can be engineered to apply dissipative forces which cool to $<$mK temperatures.  The principal difficulty with molecules, as compared to atoms, arises from the fact that their numerous internal states give rise to many leakage channels out of a cycling transition.  An excited electronic state can potentially decay into a large number of rotational and vibrational states, which stops the cooling process if there is not a laser addressing this ``dark state.''  These many states could in principle be re-pumped, like the standard hyperfine re-pumping in alkali atom magneto-optical traps (MOTs), by adding more and more lasers.  However, since typically $\gtrsim 10^4$ photons are needed to slow, capture, and cool a typical species, this could require dozens of lasers and be truly impractical for a randomly chosen molecule.  Note that this can be suppressed by relying on stimulated forces instead of spontaneous forces~\cite{Jayich2014,Kozyryev2018BiCh,Long2019,Wenz2020}, thereby reducing (though not eliminating) the role of uncontrolled spontaneous decay.
 
\begin{figure}[ht]
\centering
\includegraphics[width=0.3\textwidth]{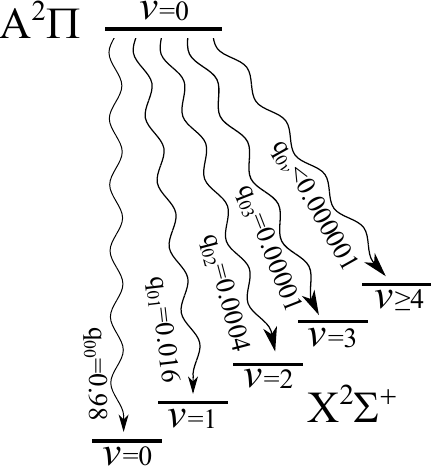}
\caption{Franck-Condon factors for SrF, the first molecule to be laser-cooled and trapped~\cite{Shuman2010,Barry2014}.  The $v=0$ state in the lowest electronic excited state decays without changing $v$ about $98\%$ of the time.  Re-pumping the $v=1,2,3$ levels enables scattering of $\gtrsim 10^6$ photons before the molecule ends up in a dark state with $v\geq 4$.}
\label{fig:FCF}
\end{figure}

One key ingredient to the solution is to find molecules unlikely to change their vibrational state during an electronic decay, of which there are actually a fair number~\cite{DiRosa2004}.  As long as the potential energy surface (PES) of the ground $\ket{g}$ and excited states $\ket{e}$ have a similar shape, the molecule is unlikely to change its vibrational state during a decay~\cite{Herzberg1989}.  This is because, to good approximation, the electronic transition is instantaneous compared to the vibrational motion of the molecule (Born-Oppenheimer approximation), so the vibrational wavefunction $\ket{v}$ is diabatically re-projected onto the eigenstates of the new PES.  This is quantified by the Franck-Condon factors (FCFs) of the electronic transition,  $q_{v'v}=|\!\braket{e,v'|g,v}\!|^2$.  If the two PESs are identical then $q_{v'v}=\delta_{v'v}$, the vibrational wavefunction will not change (within this approximation), and this transition is said to be ``diagonal'' as the array of FCFs would be a diagonal matrix.  While no transition in any molecule is ever truly diagonal, as long as it is ``nearly diagonal'' then vibrational leakage is significantly reduced, and laser cooling can occur.  For example, the first molecule to be laser-cooled and trapped, SrF~\cite{Shuman2010,Barry2014}, has $q_{00}\approx 0.98$ for the $X^2\Sigma^+\rightarrow A^2\Pi_{1/2}$ transition, and can scatter $\gtrsim 10^6$ photons with three repump lasers as shown in figure~\ref{fig:FCF}.

One case in which this condition is sometimes satisfied occurs when the valence electrons being addressed by the laser do not participate in the chemical bond, so that the electronic motion is largely decoupled from the nuclear motion and therefore the PES.  For this reason, molecules like alkaline-earth (Be, Mg, Ca, Sr, Ba, Ra) fluorides tend to work well.  The metal atom has a valence $s^2$ configuration, with one electron going to the electronegative fluorine and the other remaining centered on the metal nucleus.  This single, metal-centered $s$ electron gives rise to a $^2\Sigma$ ground state, with a $^2\Pi$ excited state often suitable for laser cooling, very analogous to the $^2S\rightarrow{}^2P$ transitions used in alkali atoms. Similarly, fluorides of the boron column (B, Al, Ga, In, Tl), which have $s^2p^1$ valence (ignoring complications from the $d$ electrons), leave a metal-centered $s^2$ configuration with laser-coolable $^1\Sigma\rightarrow{}^1\Pi$ transitions analogous to the $^1S\rightarrow{}^1P$ transitions in alkaline-earth atoms.

\begin{figure}[ht]
\centering
\includegraphics[width=0.45\textwidth]{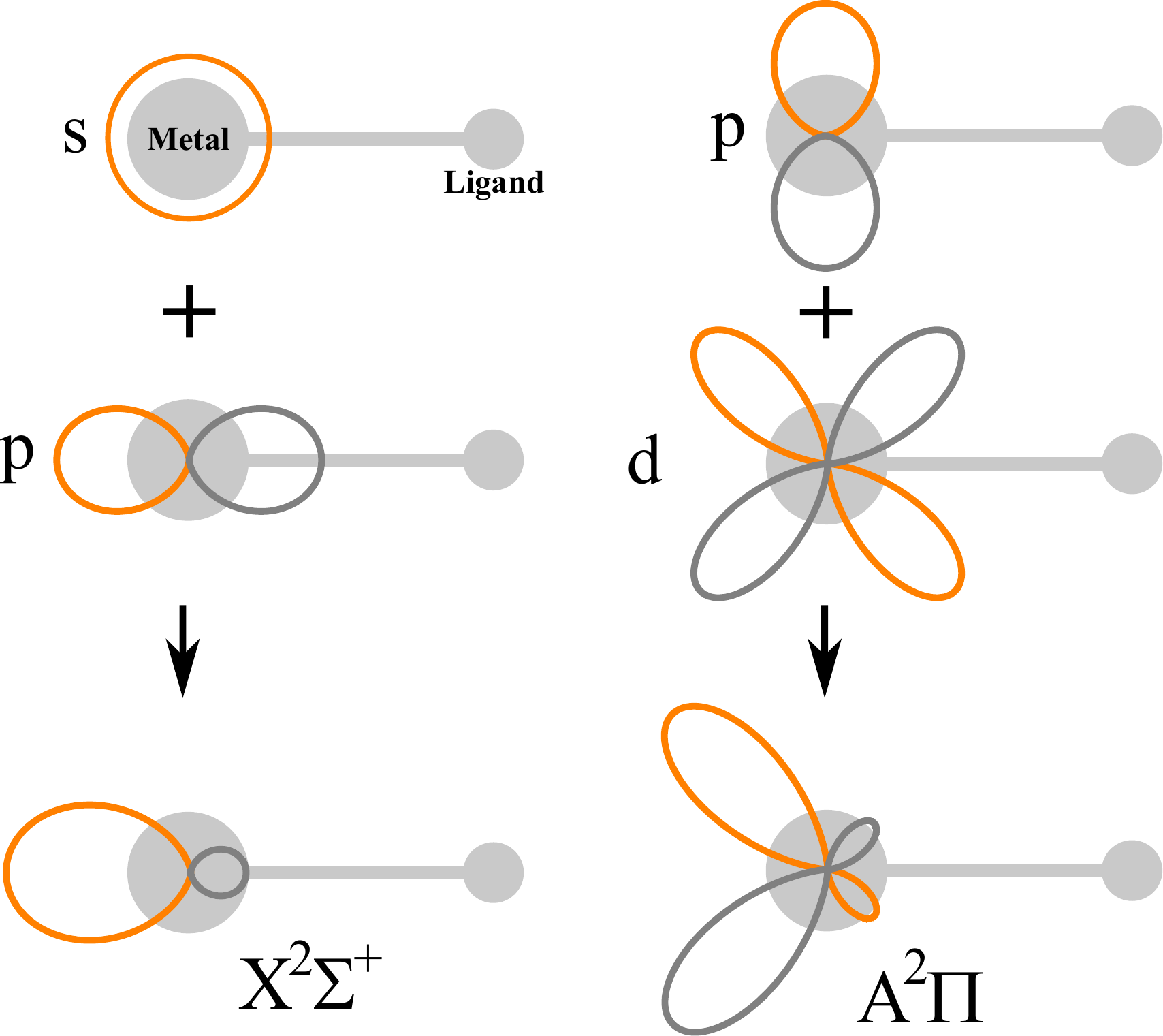}
\caption{Schematic of hybridization in alkaline-earth fluoride molecules~\cite{Ellis2001}.  The $^2\Sigma$ ground state is nominally a metal-centered $s$ electron, but hybridization with a $p$ orbital results in a wavefunction pushed away from the bonding region.  The $^2\Pi$ excited state has similar $p-d$ hybridization, so decays $A\rightsquigarrow X$ are unlikely to change the vibrational state. }
\label{fig:Hybridization}
\end{figure}

This intuitive picture is useful, but should be used with caution as these conditions are neither necessary nor sufficient.  For example, MgO is fairly diagonal, while AlO is not~\cite{Nicholls1962}, though one might guess that the reverse is the case based on the simple rules above.  Quantum chemical calculations can shed light on when and why a molecule is diagonal, and tell us that orbital hybridization is critical~\cite{Ellis2001,Isaev2016Poly,Li2019,Ivanov2019}.  In our metal-centered $s$ electron example, the chemical bond induces hybridization of the $s$ electron with the excited $p$ state, resulting in an electronic wavefunction which is pushed away from the chemical bond, as shown in figure \ref{fig:Hybridization}.  The excited $p$ state can similarly hybridize with $d$ states, and is also pushed away from the chemical bond.  This separation of electronic and vibrational degrees results in a diagonal transition.

Importantly, this discussion made no assumptions about the ligand being a single atom.  As long as the bonding partner results in a similar hybridization of orbitals, the molecule will be diagonal.  This is the case for a large number of ligands~\cite{Augenbraun2020ATM,Ellis2001,Isaev2016Poly,Li2019,Ivanov2019,Kozyryev2016Poly}, leading to a wide range of polyatomic analogues to metal fluorides, including --OH, --CCH, --NC, --CN, --CH$_3$, --OCH$_3$, --SH, --NH$_2$, and many more with increasing complexity.  This means that suitable polyatomic molecules can be laser-cooled in a manner conceptually very similar to diatomics, though with more modes to repump, and was first experimentally demonstrated in 2017 with SrOH~\cite{Kozyryev2017SrOH}.  This relative insensitivity to the ligand means that the ligand itself can contain something interesting, for example another cycling center, or an atom with some desirable property which cannot be laser-cooled itself~\cite{Puri2017,ORourke2019,Ivanov2020,Klos2020}.  This also means that the polyatomic analogue will generally maintain other useful properties of the diatomic electronic structure, such as sensitivity to the fundamental symmetry violations~\cite{Isaev2017RaOH} to be discussed later.

So far we have discussed only vibration, but rotation plays an important role in molecular laser cooling as well.  Because rotational level changes must follow selection rules due to parity and conservation of angular momentum, unlike vibrational level changes\footnote{There are selection rules regarding vibrational angular momentum, which reduces the number of states needing to be re-pumped in a linear molecule, though they are not strictly obeyed~\cite{Kinsey-Nielsen1986,Augenbraun2020YbOH,Baum2020Cycling}.}, the number of potential rotational loss channels is limited -- decays can only occur to states of opposite parity, and satisfying $\Delta J\leq 1$.  However, rotational state-changing decays can be eliminated through clever application of the selection rules with highly non-trivial consequences~\cite{Stuhl2008}.  We will not go into the details of the idea, or experimental implementation~\cite{Tarbutt2018,McCarron2018,Stuhl2008,Hummon2013,Tarbutt2015PRA,Norrgard2016,Anderegg2017}, largely because the solutions are quite general.  However, we will point out that non-linear molecules present additional challenges due to their more complex rotational structure, though these challenges can be addressed as well~\cite{Kozyryev2016Poly,Augenbraun2020ATM,Mitra2020}.

Rotational structure is also important because it necessitates efficient pre-cooling.  Thermal sources on the order of $\sim1,000$~K would disperse the molecule population over hundreds or thousands of internal states, since rotational energy scales are $\sim 1$~K.  The technique used by current molecular laser cooling experiments is the cryogenic buffer gas beam (CBGB)~\cite{Maxwell2005,Hutzler2012}, which uses an inert gas in a cryogenic environment (typically helium around $2-5$~K) to create cold, slow, and intense beams of essentially any species, including molecular radicals suitable for laser cooling.

Molecules which have been laser slowed, cooled, and trapped in 3D are limited to the diatomic species SrF~\cite{Barry2014}, CaF~\cite{Anderegg2017,Truppe2017SubDoppler}, and YO~\cite{Collopy2018}.  The diatomic molecules YbF~\cite{Lim2018} and BaH~\cite{McNally2020} have been laser-cooled, but not yet trapped.  Laser-cooled polyatomic molecules include SrOH~\cite{Kozyryev2017SrOH}, CaOH~\cite{Baum20201DMOT}, YbOH~\cite{Augenbraun2020YbOH}, and CaOCH$_3$~\cite{Mitra2020}, with a large number of proposed other species which could be cooled using these methods~\cite{Augenbraun2020ATM,Isaev2016Poly,Ivanov2019,Kozyryev2016Poly,Ivanov2020,Klos2020,Isaev2017RaOH}.  Even in the short time that these systems have been available, many advanced cooling and trapping methods have been implemented such as magnetic trapping~\cite{McCarron2018,Williams2018,Caldwell2019}, optical trapping~\cite{Anderegg2018}, application of spontaneous forces~\cite{Kozyryev2018BiCh}, tweezer trapping~\cite{Anderegg2019}, and a variety of sub-Doppler cooling methods~\cite{Kozyryev2017SrOH,Mitra2020,Truppe2017SubDoppler,Augenbraun2020YbOH,Caldwell2019,Anderegg2018,Cheuk2018,Ding2020} to reach temperatures as low as 5~$\mu$K.

There are other methods to directly cool molecules to ultracold temperatures, for example sympathetic~\cite{Lara2006,Tscherbul2011Sympathetic}, evaporative~\cite{Stuhl2012,Reens2017}, or optoelectrical~\cite{Zeppenfeld2012,Prehn2016} cooling.  These methods do not rely on the availability of closed cycling transitions, and are therefore applicable to a different set of species.  Additionally, techniques for precision measurement and control using trapped molecular ions~\cite{Zhou2020,Patterson2018,Lin2020} enable very sensitive measurements without the need for laser cooling.  Other approaches relevant for precision measurement involve formation of ultracold atoms into ultracold molecules~\cite{Ni2008}, and trapping molecules in gas matrices~\cite{Vutha2018}.

\section{Fundamental Symmetries}

We now shift to discussing our first application of laser-cooled polyatomic molecules to study fundamental physics.  We will consider two areas where polyatomic molecules offer some unique advantages -- the search for undiscovered sources of CP-violation, and the precision study of P-violation by the weak nuclear force.  Molecules are already playing an important role in studying fundamental symmetries and their violations, and we refer the reader to some recent reviews for detailed discussions~\cite{Safronova2018,DeMille2017,Chupp2019,Cairncross2019}.

\subsection{CP-violation}

The observed imbalance between matter and anti-matter in the universe cannot be explained~\cite{Dine2003}.  According to all known physics, the properties of matter and anti-matter are essentially identical, including how they could be produced after the big bang.  The absence of free anti-matter in the universe therefore suggests some undiscovered fundamental symmetry violating process.  One of the symmetries violated by this process, in addition to a preference of matter over anti-matter, is the simultaneous inversion of charge and parity, or CP~\cite{Sakharov1967}.  While this symmetry is violated in the Standard Model via the weak force CKM mechanism, its effects are sufficiently suppressed so that it cannot explain the matter-dominated universe.  There is therefore strong motivation to search for undiscovered CP-violating physics, and it is a major driver of research in many areas of physics~\cite{Safronova2018,Chupp2019,Khriplovich1997,Georgescu2020,Abe2020}.

Despite their high energy origin, these effects can manifest themselves as low energy observables.  Regular, Standard Model particles like electrons and neutrons can couple to this new physics via virtual, radiative corrections, similar to how the electron's magnetic moment is influenced by quantum electrodynamics through virtual couplings to photons and positrons.  These couplings will alter the electromagnetic properties of the particles, and imbue them with symmetry-violating electromagnetic interactions that can only arise due to exotic symmetry violating physics.

The classic example is a permanent electric dipole moment (EDM).  For a particle with spin $\vec{s}$, the EDM $\vec{d}$ must be aligned along the spin, $\vec{d}\propto\vec{s}$.  This is a consequence of the Wigner-Eckart theorem, or more intuitively, because the EDM is a vector quantity and spin is a particle's only internal vector quantum number.  This is empirically verified by quantum statistics; the indistinguishability of particles gives rise to the unique features of atomic and nuclear structure, which are a consequence of the fact that intrinsic spin is the only internal vector describing electrons and nucleons.  The very existence of a permanent EDM therefore violates a number of fundamental symmetries as $\vec{d}$ and $\vec{s}$ behave oppositely under parity (P) and time (T) reversal, and also CP by the theorem that the product CPT is always conserved~\cite{Khriplovich1997}.  This relates to our earlier discussion that permanent EDMs do not exist normally, but only arise as a result of exotic CP-violating physics.

\begin{figure}[ht]
\centering
\includegraphics[width=0.35\textwidth]{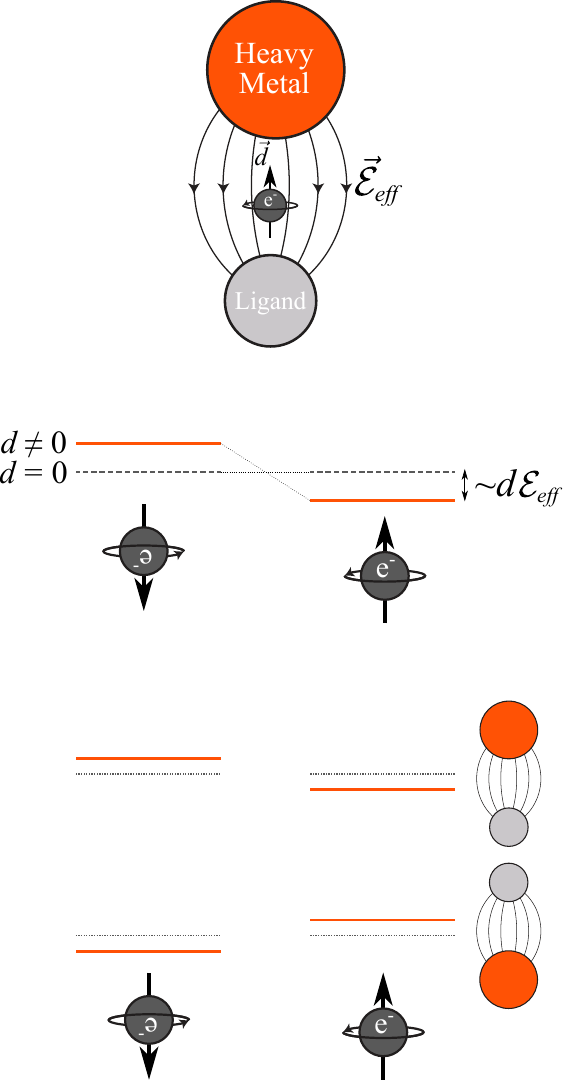}
\caption{Energy shifts due to an electron EDM, though the idea is applicable for NSM and MQM shifts as well.  \emph{Top}: The valence electrons in a molecule experience very large internal fields, especially near the nucleus of a heavy metal.  \emph{Middle}: The eEDM interacts with the internal field and gives rise to symmetry-violating energy shifts, for example between the electron spin up and down states.  \emph{Bottom}:  In a molecule with parity doublets, there are two pairs of Stark-shifted spin up/down states, one for each orientation of the fully polarized molecule.}
\label{fig:EDMShifts}
\end{figure}

The signatures of these moments can be significantly enhanced by the large internal electromagnetic fields and gradients present in atoms and molecules, especially species with a heavy nucleus, and are many orders of magnitude larger than what is achievable with laboratory fields.  The constituent electrons and nuclei experience these large fields in their bound states, giving rise to symmetry-violating energy shifts.  For example, the electron EDM gives rise to a molecular energy shift  $H_{edm}=-d_e\mathcal{E}_{ef\!f}\vec{S}\cdot\hat{n}$, where $\vec{S}$ is the effective total valence electron spin, $\Eeff$ is the effective electric field experienced by the electrons, and $\hat{n}$ is the internuclear axis, along which the internal electric field must point.  $\Eeff$ scales very rapidly with proton number, roughly as $Z^3$, and is $\sim$10-100 GV/cm for the molecules discussed here, much larger than macroscopic fields that can be realized in the lab.  This gives rise to energy shifts $\propto\pm M\propto \pm M_S$, the projection of the total angular momentum $M$, which includes the total electron spin $M_S$.  This is shown schematically in figure \ref{fig:EDMShifts}.

These energy splittings are generally too small to observe directly, for example by looking for a shift in a spectral line, so instead the Ramsey method of coherent spin precession is used~\cite{Khriplovich1997}.  By creating a superposition of the $\pm M$ levels, the phase difference between them will build up linearly in time as long as the system remains coherent, leading to a sensitivity $\propto \tau N^{1/2}$, where $\tau$ is the coherence time and $N$ is the total number of measurements.  Laser cooling offers the ability to improve coherence times by orders of magnitude compared to the current best measurements with neutral molecules~\cite{ACME2018,Hudson2011}, which have $\tau\approx 1$~ms, while maintaining large molecule numbers.  However, this is not the only method to achieve long coherence times, as they are already being realized in molecular ion trap EDM searches without laser cooling~\cite{Cairncross2017,Zhou2020}.

We can infer from the previous discussion that the species must be polarized for the EDM shifts to be non-zero.  This can be seen from the EDM Hamiltonian $\propto\vec{S}\cdot\hat{n}$; an unpolarized molecule has $\langle\hat{n}\rangle=0$ so the energy shifts will average away in the lab frame.  Electrically polarizing the species gives a non-zero average $\hat{n}$ proportional to the degree of polarization $P$ defined earlier.  This fact is the principle reason why molecules have increased sensitivity compared to atoms; the internal fields are actually very similar in both atoms and molecules, as they depend almost entirely on the short-range electronic wavefunction near the nucleus~\cite{Commins2010}.  Thus the ability to achieve $\mathcal{O}(1)$ polarization in molecules, as discussed earlier, is the source of their advantage over atoms.

Molecules with parity doublets offer the additional advantage of polarization in much smaller fields.  This simplifies engineering, but also reduces many important systematic effects which are proportional to the electric field~\cite{Vutha2010}.  Another critical feature afforded by parity doublets is the ``internal co-magnetometer,'' or ability to reverse the EDM signal without changing applied fields at all~\cite{Bickman2009}.   A fully polarized molecule with parity doublets yields two (nearly)-identical level structures with opposite Stark shifts, corresponding to the different orientations of the molecule $\hat{n}$, and therefore opposite EDM shifts, as shown in figure \ref{fig:EDMShifts}.  This gives significant robustness against systematic effects resulting from  field non-uniformity and imperfect reversals, and is a critical component of the most sensitive current eEDM experiments~\cite{Cairncross2017,ACME2018}.

Since diatomic parity doubling occurs only from $\Lambda$-doubling in states with $\Lambda>0$, this rules out the types of laser-coolable diatomic molecules discussed earlier with metal-centered $s$ electrons.  However, the parity doubling generically present in polyatomic molecules can be used to bypass these problems~\cite{Kozyryev2017PolyEDM}.  Since doubling in polyatomics can occur as a result of mechanical motion of the nuclei, as opposed to electronic motion, we can design molecules with a desirable electronic structure, such as laser cooling and good CP-violation sensitivity~\cite{Isaev2017RaOH}, and include parity doublets arising from largely independent degrees of freedom.  The bending modes in linear molecules ($\ell-$doubling) or the rotation of symmetric tops about their symmetry axis ($K-$doubling) considered earlier are present in laser-coolable molecules, and offer high polarizability and internal co-magnetometers.  Additionally, these doublets enable precision molecular ion trap measurements, and therefore opens a number of interesting species to this method~\cite{Kozyryev2017PolyEDM,Flambaum2019Schiff,Fan2020}.

\subsubsection{Electron electric dipole moment.}

Diatomic molecules are currently the most sensitive platform to search for the electron EDM, with YbF~\cite{Hudson2011}, ThO~\cite{ACME2018,Baron2014}, and HfF$^+$~\cite{Cairncross2017} surpassing the most sensitive atomic experiment performed using Tl~\cite{Regan2002}.  The current limit of $|d_e| < 1.1\times10^{-29}~e$~cm is set by the ACME experiment using ThO~\cite{ACME2018}, which is already ruling out new CP-violating particles with masses $\lesssim 10$~TeV, exceeding the reach of particle accelerators and constraining interesting new physics theories such as supersymmetry and models of baryogenesis~\cite{Safronova2018,ACME2018,Engel2013}.  Each of these diatomic experiments is also implementing improvements to continue pushing their sensitivity~\cite{Lim2018,Zhou2020,Panda2019,Ho2020,Wu2020}.  There are also a number of diatomic molecules with high sensitivity to the electron EDM which can be laser-cooled, including BaF~\cite{Hao2019}, YbF~\cite{Tarbutt2013,Lim2018,Zhou2020}, HgF~\cite{Yang2019}, and RaF~\cite{Isaev2010,GarciaRuiz2020}, all of which have the structure of a metal-centered $s$ electron giving rise to a $^2\Sigma$ state described earlier.

Polyatomic molecules with $^2\Sigma$ electronic ground states are promising candidates to combine laser cooling, large values of $\Eeff$, and parity doublets in the same system~\cite{Kozyryev2017PolyEDM}.  By combining long coherence times, large numbers, and robustness against systematics, eEDM limits can be improved by orders of magnitude, exceeding the reach of even conceivable particle accelerators.   So far, the effective electric field has been computed for many molecules of the form MOH, where M=Ca, Sr, Ba, Yb, Hg, and Ra~\cite{Isaev2017RaOH,Denis2019,Prasannaa2019,Mitra2019,Gaul2020}.  As expected, their sensitivities are comparable to their diatomic analogues.  An experimental effort to search for the electron EDM using laser-cooled YbOH molecules is underway~\cite{Kozyryev2017PolyEDM,Augenbraun2020YbOH,Nakhate2019,Steimle2019YbOH,Mengesha2020,Jadbabaie2020}, as is spectroscopic exploration of other candidate Yb-containing molecules such as YbOCH$_3$~\cite{Steimle2019YbOCH3}.

\subsubsection{Nuclear CP-violation.}

CP-violation can also occur in the nucleus, and manifest as a nuclear Schiff moment (NSM) or nuclear magnetic quadrupole moment (MQM).  Both are sensitive to physics active in the nucleus, as opposed to that which couples to the electron, and are therefore largely orthogonal and complementary to eEDM searches.  NSMs and MQMs arise, for example, from nucleon EDMs, new CP-violating nuclear forces, strong force CP-violation ($\theta$, discussed again later), and more~\cite{Safronova2018,Chupp2019,Engel2013,Ginges2004,Yamanaka2017}.

A point-like nucleus with an EDM would not interact with internal electric fields since the charged nucleus would, in steady-state, move to a position where it experienced zero average field and therefore no forces.  A nuclear dipole moment therefore appears to be unobservable in an atom or molecule, which is known as Schiff's theorem\footnote{This theorem applies to the electrons equally well, but is evaded by the fact that the electron is moving relativistically~\cite{Commins2007}.}.  However, the nucleus is not a point, and the fields are not ``screened'' perfectly, resulting in a residual electric interaction called the Schiff moment, $\vec{S}\propto\vec{I}$, and giving rise to an energy shifts $\propto\vec{S}\cdot\hat{n}\propto\vec{I}\cdot\hat{n}$, where $\vec{I}$ is the nuclear spin.  The NSM therefore behaves similarly to a nuclear dipole moment interacting with an internal electric field, but depends very strongly on the structure of the nucleus.  The effect scales roughly with proton number as $\sim Z^2$, but can be enhanced by orders of magnitude in deformed nuclei~\cite{Sushkov1985,Auerbach1996}.

The most sensitive current search for a NSM is the $^{199}$Hg vapor cell experiment~\cite{Graner2016}, and is probing the $\sim 10$~TeV scale for fundamental particles and forces in the hadronic sector, including nucleon dipole moments and CP-violating nuclear forces~\cite{Safronova2018,Engel2013}.  Other atomic searches include Ra~\cite{Parker2015,Bishof2016}, Xe~\cite{Sachdeva2019,Allmendinger2019}, and Rn~\cite{Tardiff2014}.  Molecular searches include TlF~\cite{Hunter2012,Norrgard2017,Cho1991,Clayburn2020} and RaF~\cite{Isaev2010,GarciaRuiz2020,Kudashov2014}, both of which can be laser-cooled.  RaF and Ra have a further enhancement due to an octupole $(\beta_3)$ deformation of the Ra nucleus~\cite{Auerbach1996,Gaffney2013}, leading to an intrinsic Schiff moment $\gtrsim 1000$ times larger than in spherical nuclei such as Hg.   

Another nuclear CP-violating moment is the magnetic quadrupole moment (MQM), which arises due to the same physics sources as the NSM but as a magnetic instead of electric effect~\cite{Khriplovich1997,Ginges2004,Sushkov1985,Lackenby2018} and with different relative sensitives to the sources~\cite{Flambaum2014}.  In a single particle picture, the orbiting valence nucleons in a nucleus will result in an MQM if they have EDMs, as a rotating EDM creates an MQM.  The orientation of the MQM is determined by the nuclear spin $I$, which violates CP.  The MQM will interact with valence electrons to induce an EDM in the species, which can be distinguished from the eEDM by its hyperfine dependence~\cite{Sushkov1985,Flambaum2014}.  Since the MQM is not screened, the effect is typically larger in heavy nuclei compared to the NSM.  The single-nucleon picture underestimates the MQM, which is enhanced by collective effects in quadrupole-deformed $(\beta_2)$ nuclei by around an order of magnitude~\cite{Flambaum2014,Flambaum1994}.  Because both the NSM and MQM could arise due to a large number of physics sources, measurements of these effects in multiple species complement each other~\cite{Engel2013,Flambaum2014,Chupp2015}.

The experimental procedure and requirements of NSM and MQM experiments are very similar to those for eEDM, in particular the requirement for polarization.  This means that the parity doublets of polyatomic molecules are a useful resource, and again do not in general interfere with the electronic structure requirements for good sensitivity.  NSM sensitivity has been examined for a number of species, including TlCN~\cite{Kudrin2019}, RaOH~\cite{Isaev2017RaOH}, RaOH$^+$, and ThOH$^+$~\cite{Flambaum2019Schiff}, along with MQM sensitivity for BaOH~\cite{Denis2020}, LuOH$^+$~\cite{Maison2020SearchCation}, and YbOH~\cite{Denis2020,Maison2019}, which is being pursued for an MQM search in a molecular beam~\cite{Kozyryev2017PolyEDM,Nakhate2019,Jadbabaie2020}.  All of these sensitivities, like $\Eeff$, have so far been quite similar to their fluoride analogues.

\subsection{Parity violation}

While CP-violation is yet to be observed in atoms or molecules, another important symmetry violation, parity (P), has been observed in a few~\cite{Safronova2018,Roberts2015}.  The effects of the weak nuclear force on nuclei and electrons results in symmetry-violating effects such as mixing of opposite parity states.  While these effects are extremely small, they can be distinguished from purely electromagnetic effects through their symmetry-violating characteristics.  There are a number of parity violating (PV) effects present in atoms and molecules, and several methods of measurement, though we only discuss one where molecules have a particular advantage -- nuclear spin-dependent (NSD) PV and anapole moments~\cite{Safronova2018,Roberts2015}.

The anapole moment (AM)~\cite{Zeldovich1957} is a NSD PV effect arising from $W$ and $Z$ bosons exchanging between nucleons in the nucleus and modifying the electron-nucleus interaction.  This effect is equivalent to the nucleus possessing a toroidal current distribution, like a solenoid curved into a torus, whose orientation is determined by the nuclear spin.  Since a toroidal magnetic field has no effects at long range, the AM is a contact interaction of the electrons with the nucleus, and results in the mixing of opposite parity states in species with core-penetrating $s_{1/2}$ or $p_{1/2}$ electrons.  This mixing leads to PV transition amplitudes between opposite-parity states in atoms and molecules. Only one experiment has managed to observe an AM~\cite{Wood1997}, though others are underway~\cite{Safronova2018}.  Anapole moments are an important tool for studying the weak nuclear force at low energies and in the hadronic sector~\cite{Haxton2001}, and are sensitive probes for new fundamental physics~\cite{Dzuba2017Bosons}, making complementary measurements in diverse systems very desirable.

\begin{figure}[ht]
\centering
\includegraphics[width=0.4\textwidth]{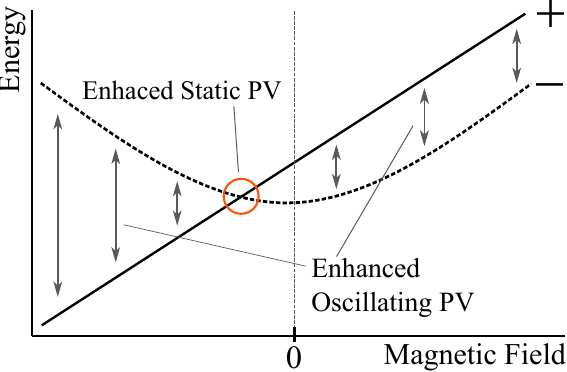}
\caption{ Enhancement of parity-violating (PV) effects in polyatomic molecules.  Parity-doubled states can have roughly quadratic or linear Zeeman shifts due to hyperfine coupling, leading to tunable energy spacings.  When these states intersect, static PV effects are enhanced~\cite{Norrgard2019}.  When the splitting is resonant with the oscillation frequency of an axion-like particle (see section \ref{sec:oscillating}) oscillating PV effects are enhanced~\cite{Stadnik2014}.  Note that there are many other levels and crossings not shown for simplicity.}
\label{fig:Crossing}
\end{figure}

This mixing of opposite-parity gives molecules a distinct advantage.  The mixing is enhanced when the states are closely-spaced, and molecules have opposite parity states with much smaller spacing compared to atoms.  In addition, these states can be pushed to near-degeneracy using magnetic fields~\cite{Flambaum1985,Kozlov1991,DeMille2008Anapole,Norrgard2019}, as shown in figure \ref{fig:Crossing}.  The PV amplitude can be measured by observing its interference with the electric dipole-allowed component of the transition due to their different interactions with applied electric fields.  This method has now been refined to the point where it has demonstrated statistical and systematic sensitivity that should enable a measurement of the $^{137}$Ba AM using the BaF molecule with $\sim10\%$ precision~\cite{Altuntas2018}.

The expanded availability of opposite-parity levels in polyatomic molecules leads to further enhancements of this method~\cite{Norrgard2019,Hao2020}.  Diatomic molecules in $^2\Sigma$ states (such as in BaF~\cite{Altuntas2018}) require large magnetic fields to push rotational states into degeneracy, which are spaced by several GHz, whereas the parity doublets in polyatomic molecules are smaller by orders of magnitude.  Molecules of the form MOH, MNC, and MCN~\cite{Norrgard2019,Hao2020} for M=Be, Mg, Ca, Sr, Ba, Yb, and Ra, for example, have parity doublets in their excited bending states which enable near-degeneracies in fields of $\lesssim$100~G, compared to thousands of G for comparable diatomics~\cite{DeMille2008Anapole}.  In addition to being technically easier, this reduces systematic effects related to the fields, as well as broadening of the transition due to field non-uniformities.

Opposite-parity states within a rotational manifold which can be pushed into degeneracy at small fields are fairly generic, and arise when the doubling is not smaller than both the fine (\emph{e.g.} spin-rotation) and hyperfine energy scales in the rotational state.  This is not too difficult to satisfy, especially since the doubling scales with rotational number.  This dependence on rotational state gives additional levels of control over where the crossing occur, and offers the opportunity to measure the effect at several different crossings for additional robustness.  Furthermore, the $^2\Sigma$ electronic states in these molecules retain the intrinsic sensitivity of their diatomic counterparts~\cite{Hao2020} and are also amenable to laser cooling.  Since the sensitivity scales with number and coherence time similar to CP-violation searches, optical trapping or molecular fountains should lead to orders-of-magnitude improvement in sensitivity.  Such an experiment is currently under development~\cite{Norrgard2019}.

This approach gives an extremely wide range of nuclei to probe, from the very light to the very heavy.  While heavy nuclei are mostly sensitive to the AM, which scales roughly as $A^{2/3}$ where $A$ is the mass number, lighter nuclei are sensitive to multiple sources of NSD PV with smaller magnitudes and are easier to benchmark against calculations.  The increased sensitivity of the polyatomic approach is a compelling way to measure these effects in a wide range of nuclei~\cite{Hao2020}, as well as search for new effects such as the weak quadrupole moment~\cite{Flambaum2017}, which would enable measurements of neutron distributions in nuclei.

\subsubsection{Parity-violating energy shifts.}

The weak nuclear force also results in energy shifts between different enantiomers of chiral molecules, though this is yet to be observed experimentally due to its extremely small size~\cite{Quack2008}.  This is an area where polyatomic molecules, specifically asymmetric top molecules, are a \emph{requirement}, since a molecule needs at least 4 atoms to have a handedness at all.  A symmetric top with a CH$_3$ group (\emph{e.g.} CaOCH$_3$) can be turned into a chiral molecule by substituting for example CHDT, and such molecules offer the possibility of laser cooling as well~\cite{Augenbraun2020ATM,Isaev2016Poly,Isaev2018}.  The recent demonstration of laser-cooled CaOCH$_3$~\cite{Mitra2020} is an important step toward potentially realizing sensitive measurements in ultracold chiral molecules.

\section{New forces and fields}

The nature of dark matter and dark energy, which together make up most of the universe~\cite{Planck2020}, is not understood.  This motivates the existence of even more new fields and forces which can couple to the Standard Model, and frequently give rise to observables to which atoms and molecules are sensitive~\cite{Safronova2018,Jansen2013,Kozlov2013Review,Hanneke2020}.  Here we shall consider two cases -- fields which change ``slowly,'' resulting in drifts of fundamental constants, or fields which oscillate on laboratory timescales, resulting in oscillations of fundamental constants or inducing transitions in atoms or molecules.

\subsection{Variation of fundamental constants}

Despite our ability to precisely measure fundamental constants~\cite{Tiesinga2020} such as the fine-structure constant $\alpha=e^2/(4\pi\epsilon_0\hbar c)=1/137.035999084(21)$ and the proton-electron mass ratio $\mu=m_p/m_e=1836.15267343(11)$, one may wonder whether they are actually constants at all.  The question is intrinsically interesting, but it also has specific motivation~\cite{Safronova2018,Jansen2013,Kozlov2013Review,Hanneke2020} as these effects arise quite frequently in extensions of the Standard Model or General Relativity.  For example, a new field to explain dark energy can result in a change of $\alpha$ in order to preserve energy conservation~\cite{Bekenstein1982}, or massive dark matter fields can couple to Standard Model parameters and result in drifts in coupling constants and masses~\cite{Stadnik2015}.  Comparison of precise, laboratory measurements of these quantities over human timescales, as well as comparisons of less precise measurements but over longer cosmological or geological timescales, yield very strong constraints on such drifts.  We will focus on laboratory measurements where polyatomic molecules could make improvements, and refer the reader to reviews for more information about the very wide variety of other approaches and motivation~\cite{Safronova2018,Jansen2013,Kozlov2013Review,Hanneke2020}.

Variations of $\alpha$ and $\mu$ are particularly relevant for atomic and molecular spectroscopy.  $\mu$ dictates the masses of the interacting bodies (including the nuclei, which are determined by $m_p$ and the neutron mass $m_n\approx m_p$), and $\alpha$ sets the strength of the electromagnetic force, both of which are responsible for essentially all spectroscopic features.  For some frequency splitting $\nu$ between two states which depends on $\alpha$ and $\mu$, it is common to define the change $\Delta\nu$ resulting from changes $\Delta\alpha$ and $\Delta\mu$ by
\[ \frac{\Delta\nu}{\nu} = K_\alpha\frac{\Delta\alpha}{\alpha}+K_\mu\frac{\Delta\mu}{\mu}, \]
where the $K$ parameters are called sensitivity coefficients.  The $\nu^{-1}$ appearing on the left-hand side of the equation suggests that the sensitivity parameters are enhanced for nearly-degenerate states, which is indeed frequently the case~\cite{Jansen2013,Kozlov2013Review,Dzuba1999,Kozlov2018}.  This is because we are defining the sensitivity parameters as a fractional change in the frequency splitting, though we must be mindful that the experimental frequencies and resolutions are also relevant~\cite{Kozlov2018}.

Electronic transition energies are $\propto\alpha^2\mu_r c^2$ where $\mu_r$ is the reduced electron-nucleus mass.  Since $\mu_r\sim m_e+\mathcal{O}(\mu^{-1}),$ the sensitivity to $\mu$ variation is suppressed by $\mu^{-1}\sim 10^{-3}$, making electronic transitions sensitive almost exclusively to $\alpha$.  Transitions in optical clocks are therefore sensitive probes for $\Delta\alpha$, and comparing two different optical frequencies at different times can be used to constrain $\dot{\alpha}/\alpha$~\cite{Rosenband2008}.  Microwave clocks relying on hyperfine structure are sensitive to both $\alpha$ and $\mu$ since the nuclear magnetic moment depends on $m_p$, while the overlap of the electronic wavefunction depends on $m_e$ and $\alpha$.  They can therefore be used to probe both $\Delta\alpha$ and $\Delta\mu$, which can be combined with $\Delta\alpha$ measurements from optical clocks to place limits on both $\dot{\alpha}/\alpha$ and $\dot{\mu}/\mu$ at the $\lesssim10^{-16}$ level~\cite{Huntemann2014,Godun2014,McGrew2019}.

While atomic frequency measurements are currently the most sensitive, molecules have a number of advantages.  First, they offer the opportunity to probe variations in $\mu$ directly, for example through rotational $(\propto \mu^{-1})$ or vibrational $(\propto \mu^{-1/2})$ constants. Second, their complex internal structure gives rise to a wide range of dependencies and energy scales, and including many ``near-degeneracies'' which significantly enhance sensitivity coefficients.  

\begin{figure}[ht]
\centering
\includegraphics[width=0.35\textwidth]{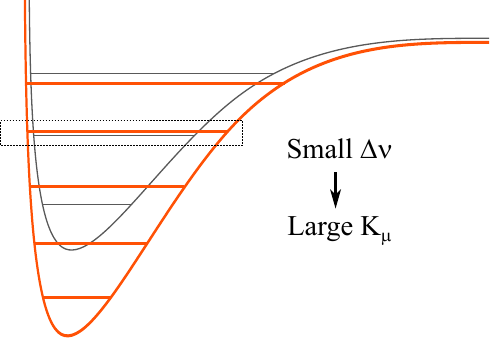}
\caption{An example of an accidental near-degeneracy in a molecule~\cite{DeMille2008Variation,Flambaum2007Variation}.  Two vibrational levels in different potential energy surfaces have a small energy splitting and different dependence on $\omega_v\propto\mu^{-1/2}$, and therefore a large enhancement factor. }
\label{fig:Degeneracy}
\end{figure}

As an example, consider a molecule with two potential energy surfaces (PESs) sharing the same dissociation asymptote, as shown in figure \ref{fig:Degeneracy}~\cite{DeMille2008Variation}. Since the PESs arises mostly from electronic effects (the chemical bond), a variation of $\mu$ can be considered as a change in the reduced mass of the oscillating bodies; in other words, the spring in the harmonic oscillator is mostly unchanged, but the mass changes, resulting in a vibration frequency change $\propto\mu^{-1/2}$.  If these two PESs have a pair nearly-degenerate vibrational levels, there can be a significant enhancement in $K_\mu$.  These degeneracies can also arise in other ways, for example from near-cancellations between spin-orbit and vibrational intervals~\cite{Flambaum2007Variation}, and are sensitive to variation of fundamental constants as long as the two states don't have the same dependence on them.  This approach was recently implemented in ultracold KRb on a transition between excited vibrational states in singlet $X^1\Sigma^+$ and triplet $a^3\Sigma^+$ electronic ground states with a sensitivity of $K_\mu\approx15,000$ to set a limit of $|\dot{\mu}/\mu|\lesssim 10^{-14}$/yr~\cite{Kobayashi2019} over a $\sim$1 year timescale.  

We shall focus on the enhancement due to near-degeneracies, though it is not the only path forward; one other approach is to use extremely sensitive measurements in diatomic molecular clocks to probe the changing vibrational energy splittings directly, without relying on accidental near-degeneracies~\cite{Safronova2018,Hanneke2020,Zelevinsky2008mu,Kondov2019}.  The complex structure of polyatomic molecules gives them many more opportunities for near-degeneracies and large sensitivity coefficients.  An example is the linear ($l$-) isomer of the C$_3$H molecule in the $^2\Pi$ ground electronic state~\cite{Kozlov2013}.  The vibrational angular momentum $\ell$ in the bending mode interacts with the electronic angular momentum to push one of resulting states down to be just $\sim$29~cm$^{-1}$ above the absolute ground state.  There are nearly-degenerate rotationally excited states in the ground and vibrating states, leading to transitions with both $K_\alpha$ and $K_\mu\sim$1,000 as the energies of the states depend on both vibrational and electronic effects.

The multiple vibrational degrees of freedom provide a more generic source of accidental degeneracy.  An important example of this is SrOH, which has nearly-degenerate vibrational levels giving significantly enhanced sensitivity $K_\mu\gtrsim1,000$~\cite{Kozyryev2018DM}.  In particular, the $(200)$ and $(03^10)$ vibrational modes in the ground $\tilde{X}^2\Sigma^+$ electronic state are within $\sim 1$~cm$^{-1}\sim30$~GHz of each other, and have rotational transitions spaced by $<200$~MHz, all of which are experimentally convenient microwave frequencies.  The sensitivity is additionally enhanced because the different vibrational modes, in particular the harmonic and anharmonic contributions, have different dependence on $\mu$.  SrOH is particularly compelling since it can be laser-cooled and trapped, and in fact was the first laser-cooled polyatomic molecule~\cite{Kozyryev2017SrOH}.  This presents an opportunity for extremely sensitive searches leveraging long coherence times and large numbers.

\subsection{Oscillating fields}\label{sec:oscillating}

We have so far focused on either static quantities like permanent, CP-violating moments, or quasi-static effects such as slowly drifting constants.  However, many proposed theories add new fields which couple to the Standard Model and result in oscillating effects.  In particular, we consider in this section oscillations whose periodicity could be observed in the laboratory, yielding different methods of experimental investigation.  

One important example is the axion, which was originally proposed~\cite{Peccei1977} as an explanation of why the strong force appears to preserve CP.  CP is not a symmetry of nature; the weak force has an order unity CP-violating phase, yet the CP-violating term $\theta$ in the strong force is constrained to be $\theta\lesssim 10^{-10}$ by neutron and $^{199}$Hg EDM experiments~\cite{Graner2016,Abel2020}.  The axion results from promoting  $\theta$ to a field which oscillates, so that it has finite amplitude but averages away.  Since $\theta$ leads to nuclear CP-violation in the form of EDMs, NSMs, and MQMs, the oscillating axion field results in oscillating (and static~\cite{Stadnik2018}) CP-violation through these channels~\cite{Stadnik2014,Graham2011,Flambaum2020SpinRotation}.  Note that these conclusions are not restricted to the axion specifically, as a generic pseudoscalar field (or ``axion-like particle'') will lead to many of the same effects.

Furthermore, axion and axion-like particles are a promising candidate for dark matter.  If dark matter is composed of such a field, we can use the local dark matter density to estimate that $\theta$ will oscillate with amplitude $\sim10^{-19}$~\cite{Graham2011}.  This may seem hopelessly small, as the limits on $\theta$ from the neutron EDM and Hg EDM experiments are around 9 orders of magnitude larger.  However, a critical feature of the axion signature is that it oscillates at a frequency $\omega_a=m_ac^2/\hbar$ given by the axion mass $m_a$.  In addition, the field is quasi-coherent; bosonic dark matter particles of mass $m\lesssim$15~eV will have overlapping de Broglie wavelengths, and can therefore be described as a classical, oscillating field~\cite{Graham2011}.  Sensitive NMR techniques may therefore be used on macroscopic samples with moles of atoms or molecules, such as in the CASPEr experiment~\cite{Graham2013,Budker2014}.

Molecules offer opportunities to search for these effects from oscillating CP-violation as well.  Existing molecular EDM data can be time-resolved to search for oscillating dipole moments up to the experimental repetition rate of $\sim1$~Hz, as was recently performed for the HfF$^+$ EDM result~\cite{Roussy2020}, so further improvements in EDM searches will advance these bounds.  Additionally, molecules offer the opportunity for resonant enhancement if the energy splitting between states can be tuned to match the axion field oscillation frequency~\cite{Flambaum2020SpinRotation}.  This results in orders-of-magnitude increases in spin precession rates, so the availability of many sources of opposite-parity states with large tunability in polyatomic molecules give them some compelling advantages.

The time-dependent parity violation of these fields also gives rise to new observables and measurement methods~\cite{Stadnik2014,Gaul2020Chiral}.  Similar to how these fields will induce oscillating CP-violation, they will also induce oscillating P-violation, for example through oscillating mixing of opposite parity states.  This mixing can be measured using qualitatively similar techniques to search for static parity violation, and therefore could benefit from the advantages of polyatomic molecules for many of the same reasons.  While static PV effects are enhanced when the states are nearly-degenerate, oscillating PV effects are resonantly enhanced by many orders of magnitude when the energy splitting between the states is equal to the frequency of the oscillating field, as shown in figure \ref{fig:Crossing}.  The abundance of opposite-parity states of polyatomic molecules with tunable splittings, combined with opportunities for laser cooling, offer exciting opportunities for future searches over a wide rage of energy scales difficult to probe with other methods.

Another approach leveraging the large span of energy scales in molecules is to search for resonant absorption of DM~\cite{Arvanitaki2018,Flambaum2020Transitions}.  The coherent, oscillating coupling to the molecular constituents can result in molecular excitation, which can then be detected from subsequent spontaneous decay or via state-specific detection of the excited state.  If the transition is tuned using the Stark or Zeeman effect to be resonant with the oscillation of the field, there will be  a resonance in the absorption rate, yielding a method to differentiate the signal from backgrounds.

\section{Outlook}

This entire perspective is an outlook, as research with laser-cooled polyatomic molecules (and even diatomic molecules) is really just getting started.  Despite being on the scene for only a few years, there are many exciting results and proposals.  As the list of species and techniques continue to grow, so too will the possibilities of what to do with these increasing complex quantum objects -- not just for precision measurement, but for the many other applications which benefit from both complexity and control, such as quantum information, simulation, sensing, chemistry, and more.

\section{Acknowledgements}

Thanks to Ben Augenbraun, Arian Jadbabaie, Ivan Kozyryev, Eric Norrgard, and Nick Pilgram for thoughts and feedback.

\section*{Appendix: Molecular structure}

Here we give a very brief overview of molecular structure in heteronuclear diatomic molecules, in particular the features relevant to the experiments discussed.  Molecular structure is an astoundingly large and complex field, and we will not even scrape the surface~\cite{Townes2012,Herzberg1967,Herzberg1989,Brown2003}.

Electronic states of molecules have term symbols $^{2S+1}\Lambda_\Omega$, very analogous to the $^{2S+1}L_J$ symbols of atoms.  In both cases $S$ denotes the total electronic spin, while $\Lambda=\vec{L}\cdot\hat{n}$ and $\Omega=\vec{J}\cdot\hat{n}$ denote the projections of the electronic orbital angular momentum $\vec{L}$ and total electronic angular momentum $\vec{J}$ onto the internuclear axis $\hat{n}$.  The strong electrostatic forces quantize the orbital angular momentum $\vec{L}$ along the internuclear axis, hence the appearance of $\Lambda$ in the term symbol.  When written in the term symbol, $\Lambda$ is denoted by the capital Greek versions of the usual atomic term symbols: $\Sigma,\Pi,\Delta,\ldots$ denote $\Lambda=0,1,2,\ldots$.  We also define $\Sigma=\vec{S}\cdot\hat{n}$, so $\Omega=\Lambda+\Sigma$.  Electronic energy scales are similar to those in atoms.

Each electronic state is assigned a letter, with $X$ indicating the ground state.  The traditional labeling scheme is that states with the same spin multiplicity as the ground state are labeled $A,B,C,\ldots$, and those with different multiplicity are labeled $a,b,c,\ldots$ in order of increasing energy, though there are frequent exceptions to this as sometimes states are discovered out of order.  The state labels are usually written with the term symbol, for example $X^2\Sigma^+$.  The $^\pm$ superscript in $\Sigma$ states indicates the parity of the lowest rotational state.  This superscript is absent for states of higher $\Lambda$ because, as we shall see, each rotational state is split into doublets of both parity in those cases.

Spin-orbit coupling splits states with the same $S$ and $\Lambda$ but different $\Omega,$ for example the $^2\Pi_{1/2}$ and $^2\Pi_{3/2}$, very similar to alkali atoms, and with a magnitude and scaling with proton number $Z$. Spin-orbit coupling also mixes states with different $\Lambda$ and $\Sigma$, but the same $\Omega$~\cite{Lefebvre2004}, meaning that $\Lambda$ and $\Sigma$ are not strictly good quantum numbers (technically, $\Omega$ isn't either~\cite{Lefebvre2004}).  However, in the limit of small spin-orbit coupling, for example in a molecule with light atoms or arising from an electronic state with $L=\Lambda=0$, these are reasonably good quantum numbers, and will work well for the molecules we consider (even the heavy ones, since we will primarily consider $\Sigma$ states.)  The different coupling schemes of all the angular momenta are described by Hund's cases, which we will not consider here, but are of critical importance to understanding the details of electronic and rotational structure~\cite{Brown2003}.

Each electronic state has a potential energy surface (PES) indicating the energy of the system as a function of internuclear separation.  Bound states posses an energy minimum at an equilibrium distance $r_e$, typically a few \AA, and a spectrum of vibrational levels corresponding to excited states of this potential.   Vibrational frequencies $\omega_v$ are typically in the range of $\sim100-1,000$~cm$^{-1}$, where the conventional unit cm$^{-1}$ means frequency of $c/(1$~cm$)\approx30$~GHz or an energy of $\approx$0.1~meV.  Vibrational frequencies are therefore in the tens to few hundred THz, or the near- to mid-infrared.  Because the PES is anharmonic, the lowest few vibrational states might have an approximately even spacing, but higher levels rapidly depart from linear spacing.  Even near the bottom of the potential, the non-linearities in the energy spacing is usually tens of GHz or more.

Each vibrational state has a series of rotational lines, along with fine and hyperfine features.  The energy of a state with $J$ rotational quanta is given by $BJ(J+1)$, where the rotational constant $B\propto 1/(\mu r_e^2)$, where $\mu$ is the reduced mass of the nuclei.  Rotational constants vary from $\sim 100$~GHz for a heavy hydride, to $\sim 1$~GHz for a heavy-heavy molecule like I$_2$.  Typical molecules used for laser cooling and for precision measurements are heavy fluorides or oxides, which have rotational constants of $\sim 10$~GHz.  The rotational constant actually depends on the vibrational level, for example because the distance between the molecules changes due to the skewed character of the PES, but that is usually not a significant effect for our purposes.

All heteronuclear molecules have the structure described so far, so further description would require additional specification of the molecular details -- and rapidly gets very complicated.  We will not need the details of all these possible interactions, save for one which is important to a number of applications discussed here.  

In states with $\Lambda=0$, the ladder of rotational states alternates parity.  In states with $\Lambda>0$, each rotational state is split into a doublet of opposite parity states through an effect called $\Lambda$-doubling.  Note that in molecules with large spin-orbit coupling where $\Lambda$ is not a good quantum number this effect is usually called $\Omega$-doubling, but the origin of the effect is the same.  States with $\pm\Lambda$ are nominally degenerate due to time-reversal symmetry.  Since $\Lambda$ is a parity-odd quantity, the field-free eigenstates are even and odd combinations of wavefunctions with defined $\Lambda$, that is, $\e^{i\Lambda\phi}\pm e^{-i\Lambda\phi}\propto\cos(\Lambda\phi),\sin(\Lambda\phi)$ where $\phi$ is the azimuthal angle about the internuclear axis.  Notice that these states constitute a doublet of opposite parity states, or a parity doublet.  These states are split due to Coriolis-type interactions between the electrons and the molecular rotation~\cite{Brown2003,Hinkley1972}, very similar to the sources of doubling discussed in the text.

\bigskip

\bigskip

\section*{References}

\bibliographystyle{iopart-num}
\bibliography{references.bib}

\end{document}